\documentclass[a4paper,12pt]{article}
\usepackage[esperanto,portuguese,english]{babel}
\usepackage{epsfig}
\usepackage{float}

\newcommand{\ppparallel}[1]{} 
\newcommand{\ppmaldekstra}[1]{} 
\newcommand{\ppdekstra}[1]{}
\usepackage{parallel} \renewcommand{\ppparallel}[1]{#1} 
\newcommand{\comentario}[1]{}

\ppparallel{    
\newlength{\pplw}\setlength{\pplw}{0.48\textwidth}
\newlength{\pprw}\setlength{\pprw}{0.50\textwidth}

\newcommand{\ppn}{\noindent}              
\newcommand{\ppl}[1]{\ParallelLText{\selectlanguage{esperanto}#1}}
\newcommand{\ppr}[1]{\ParallelRText{\selectlanguage{portuguese}#1}\ppp}
\newcommand{\ppln}[1]
{\ParallelLText{\ppn \selectlanguage{esperanto}#1}} 
\newcommand{\pprn}[1]
{\ParallelRText{\ppn \selectlanguage{portuguese}#1}\ppp} 

\newcommand{\ppsection}[3][0ex]{\vspace{2em} 
\ppl{\section{#2} \vspace{#1}} \ppa \nopagebreak
\ppR{\section{#3}} \ppp \nopagebreak} 

\newcommand{\bea}{\vspace{-1ex}\begin{eqnarray}}
\newcommand{\eea}{\end{eqnarray}}
} 

\ppmaldekstra{
\newcommand{\ppl}[1]{\selectlanguage{esperanto}#1}
\newcommand{\ppln}[1]{\noindent \selectlanguage{esperanto}#1}
\newcommand{\ppr}[1]{\selectlanguage{portuguese}}
\newcommand{\pprn}[1]{\noindent \selectlanguage{portuguese}}
\newcommand{\ppsection}[3][0ex]{\section{#2}}

\newcommand{\bea}{\begin{eqnarray}}
\newcommand{\eea}{\end{eqnarray}}
}

\ppdekstra{
\newcommand{\ppl}[1]{\selectlanguage{esperanto}}
\newcommand{\ppln}[1]{\noindent \selectlanguage{esperanto}}
\newcommand{\ppr}[1]{\selectlanguage{portuguese}#1}
\newcommand{\pprn}[1]{\noindent \selectlanguage{portuguese}#1}
\newcommand{\ppsection}[3][0ex]{\section{#3}}

\newcommand{\bea}{\begin{eqnarray}}
\newcommand{\eea}{\end{eqnarray}}

}

\title{{\bf Algorithm for structure constants \ppparallel{\\ Algoritmo por konstantoj de strukturo}}}
\author{F.M. Paiva \\ 
{\small Departamento de F\'isica, Unidade Humait\'a II, Col\'egio Pedro II} \\
{\small Rua Humait\'a 80, 22261-040  Rio de Janeiro-RJ, Brasil; fmpaiva@cbpf.br} 
\vspace{.7ex} \\
A.F.F. Teixeira \\
{\small Centro Brasileiro de Pesquisas F\'isicas} \\
{\small 22290-180 Rio de Janeiro-RJ, Brasil; teixeira@cbpf.br}}

\begin{document}
\selectlanguage{esperanto}
\maketitle
\thispagestyle{empty}

\begin{abstract}\selectlanguage{english}
In a $n$-dimensional Lie algebra, random numerical values are assigned by computer to $n(n-1)$ especially selected structure constants. An algorithm is then created, which calculates without ambiguity the remaining constants obeying the Jacobi conditions. Differently from others, this algorithm is suitable even for poor personal computer.    

\ppparallel{\selectlanguage{esperanto} En $n$-dimensia algebro de Lie, hazardaj numeraj valoroj estas asignitaj per komputilo al $n(n-1)$ speciale elektitaj konstantoj de strukturo. Tiam algoritmo estas kreita, kalkulante senambigue la ceterajn konstantojn obeante kondi\^cojn de Jacobi. Malsimile al aliaj algoritmoj, tiu \^ci ta\u ugas e\^c por malpotenca komputilo.}
\end{abstract}

\ppparallel{
\begin{Parallel}[v]{\pplw}{\pprw}
}

\ppparallel{\section*{\vspace{-2em}}\vspace{-2ex}}   

\ppsection[0.6ex]{Enkonduko}{Introduction}                                        \label{Sek1}

\ppln{Bonkonate, $\!$ bazaj $\!$ vektoroj $\!$ $e_i$ $\!$  $\!$ de $n$-di\-mensia algebro de Lie obeas regulon~\cite[pa\^go 383]{Cornwell}}
\pprn{As is well known, the basis vectors $e_i$ of a $n-$dimensional Lie algebra obey rule~ \cite[page 383]{Cornwell}} 

\bea                                                                               \label{comut}
[e_i, e_j]_{_{-}}={C_{ij}}^k e_k\,,  
\eea 

\ppln{kie indicoj varias de 1 al $n$, kaj kie konstantoj de strukturo ${C_{ij}}^k$ estas antisimetriaj en malsupraj indicoj: ${C_{ij}}^k=-{C_{ji}}^k$. Ni profitas tiun antisimetrion por skribi nur konstantojn ${C_{ij}}^k$ havante\, $i<j$. Plue, $\!$ ili $\!$ devas $\!$ obei $\!$ kondi\^cojn de Jacobi} 
\pprn{where the indices vary from 1 to $n$, and where the constants of structure ${C_{ij}}^k$ are antisymmetric in the lower indices: ${C_{ij}}^k=-{C_{ji}}^k$. We profit from that antisymmetry to write only constants ${C_{ij}}^k$ having $i<j$. Still, they must obey the Jacobi conditions}

\bea                                                                                \label{Jac1}
{C_{ij}}^l{C_{kl}}^m+{C_{jk}}^l{C_{ij}}^m+{C_{ki}}^l{C_{jl}}^m=0\,.
\eea

\ppln{Tiuj kondi\^coj estas antisimetriaj en la 3 indicoj $i,j,k$, do ili agas se nur $n\geq3$.}
\pprn{These conditions are antisymmetric in the 3 indices $i,j,k$, so they act only if $n\geq3$.}

\ppl{Se $n=1$, tiam la algebro havas nur 1 bazan vektoron, kaj ^gia nura konstanto de strukturo estus ${C_{11}}^1$, nula.}
\ppr{If $n=1$, the algebra has only 1 basis vector, and its only structure constant would be ${C_{11}}^1$, null.} 

\ppl{Se $n=2$, tiam 2 konstantoj povas ekzisti, sendependaj kaj ne-nulaj: ${C_{12}}^1$ kaj ${C_{12}}^2$.}
\ppr{If $n=2$, two structure constants can occur, independent and non-null: ${C_{12}}^1$ and ${C_{12}}^2$.} 

\ppl{Se $n=3$, tiam  9 ne-nulaj konstantoj  povas ekzisti: ${C_{12}}^m$, ${C_{13}}^m$, ${C_{23}}^m$, kun $m=1,2,3$\,. Tamen la 3 kondi\^coj de Jacobi}
\ppr{If $n=3$, nine non-null structure constants can occur: ${C_{12}}^m$, ${C_{13}}^m$, ${C_{23}}^m$, with $m=1,2,3$\,. But the 3 Jacobi conditions} 

\bea                                                                                \label{cond}
{C_{12}}^k{C_{3k}}^m+{C_{23}}^k{C_{1k}}^m+{C_{31}}^k{C_{2k}}^m=0\,, \hspace{3mm} m=1,2,3 
\eea 

\ppln{reduktas tiujn 9 konstantojn al nur 6 sendependaj.}
\pprn{reduce these 9 constants to only 6 independent.} 

\ppl{Se $n=4$, tiam 24 ne-nulaj konstantoj de strukturo povas ekzisti: ${C_{12}}^m$, ${C_{13}}^m$, ${C_{14}}^m$, ${C_{23}}^m$, ${C_{24}}^m$, ${C_{34}}^m$, estante $m=1,2,3,4$. Kaj la kondi\^coj de Jacobi estas 16:}
\ppr{If $n=4$ then 24 non-null structure constants can exist: ${C_{12}}^m$, ${C_{13}}^m$, ${C_{14}}^m$, ${C_{23}}^m$, ${C_{24}}^m$, ${C_{34}}^m$, with $m=1,2,3,4$. And the Jacobi conditions are 16:} 

\bea                                                                                \label{Jac2}
{C_{12}}^k{C_{3k}}^m+{C_{23}}^k{C_{1k}}^m+{C_{31}}^k{C_{2k}}^m&=&0\,, \\
\nonumber {C_{12}}^k{C_{4k}}^m+{C_{24}}^k{C_{1k}}^m+{C_{41}}^k{C_{2k}}^m&=&0\,, \\
\nonumber {C_{13}}^k{C_{4k}}^m+{C_{34}}^k{C_{1k}}^m+{C_{41}}^k{C_{3k}}^m&=&0\,, \\
\nonumber {C_{23}}^k{C_{4k}}^m+{C_{34}}^k{C_{2k}}^m+{C_{42}}^k{C_{3k}}^m&=&0\,,
\eea

\ppln{kun $m=1,2,3,4$. Ni povus pensi, ke tiuj 16 kondi\^coj reduktas la 24 konstantojn de strukturo al nur $24 - 16 = 8$ sendependaj, sed tio ne veri\^gas. Fakte, nur 12 el tiuj kondi\^coj estas sendependaj, tiom reduktante de 24 al $24 - 12 = 12$ la nombron de sendependaj konstantoj de strukturo.}
\pprn{with $m=1,2,3,4$. We could think that these 16 conditions reduce the 24 structure constants to only $24 - 16 = 8$ independent, but that is not true. In fact, only 12 of these conditions are independent, so reducing from 24 to $24 - 12 = 12$ the number of independent structure constants.}

\ppl{Tiu \^ci artikolo montras, ke en $n$-dimensia algebro de Lie, la plejgranda nombro de konstantoj de strukturo kies numerajn valorojn oni povas hazarde elekti estas $n(n-1)$. Por tio, ni prezentas algoritmon malkovrante neambigue la pluajn $\frac{1}{2}n(n-1)(n-2)$ konstantojn de strukturo, kondi\^ce ke tiuj $n(n-1)$ konstantoj estas konvene elektitaj.}
\ppr{This article shows that, in a $n$-dimensional Lie algebra, the maximum number of structure constants whose numerical values one can randomly choose is $n(n-1)$. To that end, we present an algorithm that uncovers unambiguously the remaining $\frac{1}{2}n(n-1)(n-2)$ structure constants, provided these $n(n-1)$ constants are conveniently chosen.}

\ppl{Anta\u ue verki la okazon de arbitra $n$, la sekvanta sekcio analizas detale la okazon $n=4$. Tiu analizo montras kiel sinsekvaj sistemoj de $n$ linearaj ekvacioj por $n$ variabloj aperas kaj estas solvitaj, la\u ulonge la algoritmo.}
\ppr{Before working the case of arbitrary $n$, the next section analyses with detail the case $n=4$. That analysis shows how consecutive systems of $n$ linear equations for $n$ variables appear and are solved, in the course of the algorithm.}

\ppsection[0.6ex]{Se $n=4$}{If $n=4$}                                         \label{kvardim}

\ppln{Ni interkonsentas la jenan notacion por kon\-di\^coj de Jacobi:}
\pprn{We agree the following notation for Jacobi conditions:}

\bea                                                                              \label{Jacobi}
{J_{abc}}^m:=\left({C_{ab}}^k{C_{ck}}^m+{C_{bc}}^k{C_{ak}}^m+{C_{ca}}^k{C_{bk}}^m=0\right)\,; 
\eea

\ppln{kaj ni profitas ilian antisimetrion por uzi nur ekvaciojn ${J_{abc}}^m$ kun $a<b<c$.}
\pprn{and we profit their antisymmetry to use only the equations ${J_{abc}}^m$ with $a<b<c$.}

\ppl{Ni vidu kio okazas en $4$-dimensia algebro de Lie se ni faras komputilo hazarde elekti numerajn valorojn por konstantoj de strukturo havante malsupran indicon 1. Tiam la 12 konstantoj ${C_{1j}}^k$ kun $j=2,3,4$ kaj $k=1,2,3,4$ estos numeroj, ekde tie \^ci.}
\ppr{Let us see what happens in a $4$-dimensional Lie algebra if we make a computer choose random numerical values for the structure constants with one lower indice 1. Then the 12 constants ${C_{1j}}^k$ with $j=2,3,4$ and $k=1,2,3,4$ will be numbers, from here on.}

\ppl{Skribu la 4 ekvaciojn ${J_{123}}^m$:}
\ppr{Write the 4 equations ${J_{123}}^m$:} 

\bea                                                                                \label{J123}
{J_{123}}^1:\hspace{5mm} \alpha_0+\alpha_m{C_{23}}^m+\alpha_5{C_{24}}^1+\alpha_6{C_{34}}^1&=&0, \\
\nonumber {J_{123}}^2:\hspace{6mm}  \beta_0+\beta_m{C_{23}}^m+\beta_5{C_{24}}^2+\beta_6{C_{34}}^2&=&0,\\
\nonumber {J_{123}}^3:\hspace{7mm} \gamma_0+\gamma_m{C_{23}}^m+\gamma_5{C_{24}}^3+\gamma_6{C_{34}}^3&=&0, \\
\nonumber {J_{123}}^4:\hspace{8mm} \delta_0+\delta_m{C_{23}}^m+\delta_5{C_{24}}^4+\delta_6{C_{34}}^4&=&0,
\eea

\ppln{kie $m=1,2,3,4$ kaj grekaj simboloj estas numeroj. Ni rimarkas, ke tiuj 4 ekvacioj estas linearaj por \^ciuj konstantoj ${C_{ij}}^k$ kun \mbox{$i\neq1$}. Ni solvas tiujn 4 ekvaciojn ${J_{123}}^m$ por la 4 konstantoj ${C_{23}}^m$. Tiuj \^ci fari\^gas linearaj kombinoj de la 8 konstantoj ${C_{24}}^m$ kaj ${C_{34}}^m$:}
\pprn{where $m=1,2,3,4$ and greek symbols are numbers. We note that these 4 equations are linear for all constants ${C_{ij}}^k$ with \mbox{$i\neq1$}. We solve these 4 equations ${J_{123}}^m$ for the 4 constants ${C_{23}}^m$. These become linear combinations of the 8 constants ${C_{24}}^m$ and ${C_{34}}^m$:}

\bea                                                                                 \label{C23}
{C_{23}}^1&=&\epsilon_0+\epsilon_m{C_{24}}^m+\epsilon_{m+4}{C_{34}}^m, \\
\nonumber {C_{23}}^2&=&\eta_0+\eta_m{C_{24}}^m+\eta_{m+4}{C_{34}}^m, \\
\nonumber {C_{23}}^3&=&\,\iota_0+\,\iota_m{C_{24}}^m+\,\iota_{m+4}{C_{34}}^m, \\
\nonumber {C_{23}}^4&=&\kappa_0+\kappa_m{C_{24}}^m+\kappa_{m+4}{C_{34}}^m, 
\eea 

\ppln{kie $m=1,2,3,4$ kaj grekaj simboloj estas numeroj. Ni metas esprimojn (\ref{C23}) de ${C_{23}}^m$ en la 4 ekvacioj ${J_{124}}^m$, kaj refoje ricevas linearecon por ${C_{24}}^p$ kaj ${C_{34}}^q$:}
\pprn{where $m=1,2,3,4$ and greek symbols are numbers. We insert the expressions (\ref{C23}) of ${C_{23}}^m$ in the 4 equations ${J_{124}}^m$, and again obtain linearity for ${C_{24}}^p$ and ${C_{34}}^q$:}

\bea                                                                                \label{J124}
{J_{124}}^1: \hspace{5mm} \lambda_0+\lambda_m{C_{24}}^m+\lambda_{m+4}{C_{34}}^m&=&0, \\
\nonumber {J_{124}}^2: \hspace{5mm} \mu_0+\mu_m{C_{24}}^m+\mu_{m+4}{C_{34}}^m&=&0, \\
\nonumber {J_{124}}^3: \hspace{7mm} \nu_0+\nu_m{C_{24}}^m+\nu_{m+4}{C_{34}}^m&=&0, \\ 
\nonumber {J_{124}}^4: \hspace{6mm} \pi_0+\pi_m{C_{24}}^m+\pi_{m+4}{C_{34}}^m&=&0,
\eea

\ppln{kie $m=1,2,3,4$ kaj la grekaj simboloj estas numeroj. Ni solvas tiujn 4 ekvaciojn por ricevi la 4 variablojn ${C_{24}}^m$, kiuj fari\^gas linearaj kombinoj de la 4 variabloj ${C_{34}}^m$:}
\pprn{where $m=1,2,3,4$ and the greek symbols are numbers. We solve these 4 equations for the 4 variables ${C_{24}}^m$, that become linear combinations of the 4 variables ${C_{34}}^m$:}

\bea                                                                                 \label{C24}
{C_{24}}^1=\rho_0+\rho_m{C_{34}}^m&,&\hspace{5mm} {C_{24}}^2=\sigma_0+\sigma_m{C_{34}}^m, \\
\nonumber {C_{24}}^3=\tau_0+\tau_m{C_{34}}^m&,& \hspace{5mm} {C_{24}}^4=\phi_0+\phi_m{C_{34}}^m,
\eea

\ppln{kie $m=1,2,3,4$ kaj la grekaj simboloj estas numeroj. Fine, ni metas tiujn kombinojn de ${C_{34}}^m$ en la 4 kondi\^coj de Jacobi ${J_{134}}^m$, kaj solvas la linearajn ekvaciojn. Tiel ni ricevas la numerajn valorojn de la 4 konstantoj ${C_{34}}^m$. Sekve, ni returne kalkulas la valorojn de ${C_{24}}^m$ uzante (\ref{C24}), kaj de ${C_{23}}^m$ uzante (\ref{C23}).}
\pprn{where $m=1,2,3,4$ and the greek symbols are numbers. Finally, we insert these combinations of ${C_{34}}^m$ into the 4 Jacobi conditions ${J_{134}}^m$, and solve the linear equations. We thus obtain the numerical values of the 4 constants ${C_{34}}^m$. Then we calculate back the values of ${C_{24}}^m$ using (\ref{C24}), and of ${C_{23}}^m$ using (\ref{C23}).}

\ppl{Oni vidas, ke la serio de la algoritmo por $n=4$ estis}
\ppr{One sees that the sequence of the algorithm for $n=4$ was}

\bea 
\{{C_{1j}}^m\} \rightarrow{J_{123}}^m \rightarrow {C_{23}}^m \rightarrow {J_{124}}^m \rightarrow {C_{24}}^m \rightarrow {J_{134}}^m \rightarrow {C_{34}}^m\,,  
\eea

\ppln{kaj la serio de finaj kalkuloj de dependaj konstantoj de strukturo estis}
\pprn{and the sequence of final calculation of dependent structure constants was}

\bea                                                                                            
{C_{34}}^m \rightarrow {C_{24}}^m \rightarrow {C_{23}}^m\,. 
\eea 

\ppl{Se ni metas la 12 numerajn valorojn asignitaj por konstantoj ${C_{1j}}^m$, kaj la 12 numerajn valorojn ricevitajn por la aliaj ${C_{ij}}^m$, en iu ajn el la 4 kondi\^coj de Jacobi ne uzitaj, ${J_{234}}^m$, ni ricevos $0=0$.}
\ppr{If we insert the 12 numerical values assigned to constants ${C_{1j}}^m$, and the 12 numerical values obtained for the other ${C_{ij}}^m$, into any of the 4 Jacobi conditions not used, ${J_{234}}^m$, we shall obtain $0=0$.} 

\ppsection[0.6ex]{Por arbitra $n$}{For arbitrary $n$}                             \label{ndim}

\ppln{Kondi\^coj de Jacobi (\ref{Jacobi}) estas kvadrataj en konstantoj de strukturo. Tial, ricevi numerajn valorojn de konstantoj de $n$-dimensia algebro de Lie, ekde numeraj valoroj de kelkaj konstantoj hazarde elektitaj, estas longtempa verko, ordinare. Tamen, anta\u ua sekcio sugestas simplan kaj efikan algoritmon por, poste konvena elekto de preciza nombro de konstantoj, ricevi la ceterajn konstantojn.}
\pprn{The Jacobi conditions (\ref{Jacobi}) are quadratic in the structure constants. So, obtaining numerical values for constants of a $n$-dimensional Lie algebra, out from numerical values of some constants randomly chosen, is a time consuming task, generally. However, the preceding section suggests a simple and efficient algorithm for obtaining the remaining constants, after a convenient choice of a precise number of constants.} 

\ppl{En tiu algoritmo, unue ni aran\^gas la konstantojn de strukturo ${C_{ab}}^m$ la\u u ordo pligrandi\^ganta en la malsupraj indicoj:}
\ppr{In that algorithm, we first arrange the structure constants ${C_{ab}}^m$ in increasing order of lower indices:} 

\bea                                                                                            
{C_{12}}^m,\,{C_{13}}^m,\,...\,,\,{C_{1n}}^m,\,{C_{23}}^m,\,{C_{24}}^m,\,...\,,\,{C_{2n}}^m,......\,,\,{C_{(n-1)n}}^m\,, 
\eea

\ppln{kaj simile aran\^gas la kondi\^cojn de Jacobi ${J_{1bc}}^m$:}
\pprn{and similarly arrange the Jacobi conditions ${J_{1bc}}^m$:}

\bea                                                                                            
{J_{123}}^m,\,{J_{124}}^m,\,...\,,\,{J_{12n}}^m,\,{J_{134}}^m,\,...\,,{J_{13n}}^m,\,......\,,\,{J_{1(n-1)n}}^m. 
\eea

\ppl{Ekde tie \^ci, la algoritmo procedas kiel en anta\u ua sekcio: unue ni faras komputilo asigni hazardajn numerajn valorojn por la $n(n-1)$ konstantoj de strukturo ${C_{1b}}^m$, kaj metas tiujn valorojn en la $n$ ekvacioj de Jacobi ${J_{123}}^m$. Poste ni solvas tiujn linearajn ekvaciojn por la $n$ variabloj ${C_{23}}^m$, kaj tiel pluen.}
\ppr{From here on, the algorithm proceeds as the preceding section: we first make a computer assign random numerical values for the $n(n-1)$ structure constants ${C_{1b}}^m$, and set these values into the $n$ Jacobi equations ${J_{123}}^m$. Then solve these linear equations for $n$ variables ${C_{23}}^m$, and so on.}

\ppl{Do la serio de la algoritmo estas}
\ppr{The sequence of the algorithm then is}

\bea \nonumber
\{{C_{1b}}^m\} \rightarrow {J_{123}}^m \rightarrow {C_{23}}^m \rightarrow {J_{124}}^m \rightarrow {C_{24}}^m \rightarrow ... \rightarrow {J_{12n}}^m \rightarrow {C_{2n}}^m \rightarrow {J_{134}}^m \rightarrow {C_{34}}^m \rightarrow 
\eea 
\vspace{-14mm} 
\bea
{}                                                                                              
\eea 
\vspace{-13mm}                                                        
\bea \nonumber  
\rightarrow{J_{135}}^m \rightarrow {C_{35}}^m \rightarrow ... \rightarrow {J_{13n}}^m \rightarrow {C_{3n}}^m \rightarrow ... \hspace{2mm} ... \rightarrow {J_{1(n-1)n}}^m \rightarrow {C_{(n-1)n}}^m\,. 
\eea

\ppl{Havante la numerajn valorojn de konstantoj ${C_{(n-1)n}}^m$, ni sekve kalkulas la valorojn de ceteraj konstantoj en inversa ordo:}
\ppr{Having the numerical values of the constants ${C_{(n-1)n}}^m$, we sequentially calculate the values of the remaining constants in the reverse order:} 

\bea                                                                                            
{C_{23}}^m \leftarrow ... \leftarrow {C_{(n-3)(n-1)}}^m \leftarrow {C_{(n-3)n}}^m \leftarrow {C_{(n-2)(n-1)}}^m \leftarrow {C_{(n-2)n}}^m \leftarrow {C_{(n-1)n}}^m. 
\eea 

\ppl{Por kontroli, ni metas valorojn de la sendependaj kaj dependaj konstantoj en iu ajn ekvacio de Jacobi ${J_{abc}}^m$ kun $a\neq1$; ni devas ricevi $0=0$.}
\ppr{To check, we introduce the values of the independent and dependent constants into any Jacobi equation ${J_{abc}}^m$ with $a\neq1$; we should obtain $0=0$.} 

\ppl{Tiu \^ci algoritmo evidentigas, ke la maksimuma nombro de sendependaj konstantoj de strukturo de $n$-dimensia algebro de Lie, kaj la responda nombro de dependaj konstantoj, estas}
\ppr{This algorithm makes evident that the maximum number of independent structure constants of a $n$-dimensional Lie algebra, and the corresponding number of dependent constants, are}

\bea                                                                                            
n(n-1)\,, \hspace{5mm} n{C_n}^2-n(n-1)=\frac{1}{2}n(n-1)(n-2)\,, 
\eea 

\ppln{respektive. Anka\u u, la nombro de sendependaj kaj de dependaj kondi\^coj de Jacobi estas, respektive,}
\pprn{respectively. Also, the number of independent Jacobi conditions and that of dependent ones are, respectively,} 

\bea                                                                                            
\frac{1}{2}n(n-1)(n-2)\,, \hspace{5mm} n{C_n}^3-\frac{1}{2}n(n-1)(n-2)=\frac{1}{6}n(n-1)(n-2)(n-3)\,.
\eea 

\ppsection[0.6ex]{Komentoj}{Comments}                                            \label{comen}

\ppln{La algoritmo prezentita tie \^ci solvas sinsekve, $\frac{1}{2}(n-1)(n-2)$ foje, sistemojn de $n$ linearaj ekvacioj por $n$ variabloj. Tiu algoritmo multe ta\u ugas al persona komputilo, e\^c de malgranda kapablo.}
\pprn{The algorithm presented here solves sequentially, $\frac{1}{2}(n-1)(n-2)$ times, systems of $n$ linear equations for $n$ variables. That algorithm is suitable for a personal computer, even of short power.}

\ppl{Vere, estas maniero matematike pli rektmetoda por ricevi la konstantojn de strukturo pendantaj de la $n(n-1)$ konstantoj, elektitaj kiel ni faris tie \^ci. Tiu alia maniero profitas, ke \^ciuj $\frac{1}{2}n(n-1)(n-2)$ ekvacioj ${J_{1bc}}^m$ estas linearaj por \^ciuj $\frac{1}{2}n(n-1)(n-2)$ konstantoj ${C_{ab}}^r$ kun $a>1$. Do, se la $n(n-1)$ konstantoj ${C_{1b}}^r$ estas donitaj, tiu alia maniero bezonas solvi nur 1 sistemon de $\frac{1}{2}n(n-1)(n-2)$ linearaj ekvacioj. Malfeli^ce, se $n$ estas granda, tiu elefanta kalkula\^{\j}o ordinare estas pluen la kapablo de disponebla persona komputilo.}
\ppr{As a matter of fact, there is a mathematically simpler way to obtain the structure constants depending on the $n(n-1)$ constants, chosen as we did here. This alternative way profits from all $\frac{1}{2}n(n-1)(n-2)$ equations ${J_{1bc}}^m$ being linear for all $\frac{1}{2}n(n-1)(n-2)$ constants ${C_{ab}}^r$ with $a>1$. So, if the $n(n-1)$ constants ${C_{1b}}^r$ are given, this alternative way simply needs solving just 1 system of $\frac{1}{2}n(n-1)(n-2)$ linear equations. Unfortunately, if $n$ is large, this elephantine calculation is generally beyond the capacity of a personal computer at disposal.}        

\selectlanguage{esperanto}
\vspace{1cm}

\end{Parallel}

\end{document}